\documentclass[11pt,twoside]{article}
\usepackage{asp2010}

\resetcounters

\begin{document}

\title{Models of Stars, Brown Dwarfs and Exoplanets} 
\author{Allard, F., Homeier, D., Freytag, B.} 
\affil{CRAL, UMR 5574, CNRS,
  Universit\'e de Lyon, 
 \'Ecole Normale Sup\'erieure de Lyon, 
 46 All\'ee d'Italie, F-69364
 Lyon Cedex 07, France}

\begin{abstract} Within the next few years, GAIA and several
instruments aiming at imaging extrasolar planets will see first light.
In parallel, low mass planets are being searched around red dwarfs which
offer more favourable conditions, both for radial velocity detection
and transit studies, than solar-type stars.  
Authors of the model atmosphere code which has
allowed the detection of water vapour in the atmosphere of Hot
Jupiters review recent advancement in modelling the stellar to
substellar transition. The revised solar oxygen abundances and cloud
model allow for the first time to reproduce the photometric and
spectroscopic properties of this transition.  Also presented are 
highlight results of a model atmosphere grid for stars, brown
dwarfs and extrasolar planets. 
\end{abstract}

\section{Introduction}

Since spectroscopic observations of very low mass
stars (late 80s), brown dwarfs (mid 90s) and extrasolar planets
(mid 2000s) are available, one of the most important challenges in
modelling their atmospheres and spectroscopic properties lies in high
temperature molecular opacities and cloud formation. 
K dwarfs show the onset of formation metal hydrides (starting
around $T_\mathrm{eff} \sim 4500$\,K), TiO and CO (below
$T_\mathrm{eff} \sim 4000$\,K), while water vapour forms in early M
dwarfs ($T_\mathrm{eff}\sim 3900-2000$\,K), and methane, ammonia and
carbon dioxide are detected in late-type brown dwarfs
($T_\mathrm{eff}\sim 300-1600$\,K) and in extrasolar giant planets. 
The latter are either observed by transit ($T_\mathrm{eff}\sim 1000-2000$\,K
depending on the spectral type of the central star and the distance to
the star) or by imaging (young planets of $T_\mathrm{eff}\sim
300-2000$\,K depending on their mass and age). 

The modelling of the atmospheres of very low mass stars (hereafter
VLMs) has evolved (as here illustrated with the development of the
\texttt{PHOENIX} atmosphere code, which has allowed the detection of
water vapour in extrasolar planets' atmospheres by Barman et al.\
2007, 2008\nocite{Barman2007,Barman2008}) with the extension of computing 
capacities from an analytical treatment of the transfer equation using
moments of the radiation field \cite[]{AllardPhDT90}, to a
line-by-line opacity sampling in spherical symmetry
\cite[]{ARAA97,NGa,NGb}, and more recently to 3D radiation transfer
\cite[]{SHB2010}. In parallel to detailed radiative transfer in an
assumed static environment, hydrodynamical simulations have been
developed to reach a realistic representation of the granulation and
its induced line shifts for the sun and sun-like stars
\citep[see][]{Freytag2011} by using a non-grey
(multi-group binning of opacities) radiative transfer with a pure
blackbody source function (scattering is neglected).

\begin{figure}[!ht]
  \plotfiddle{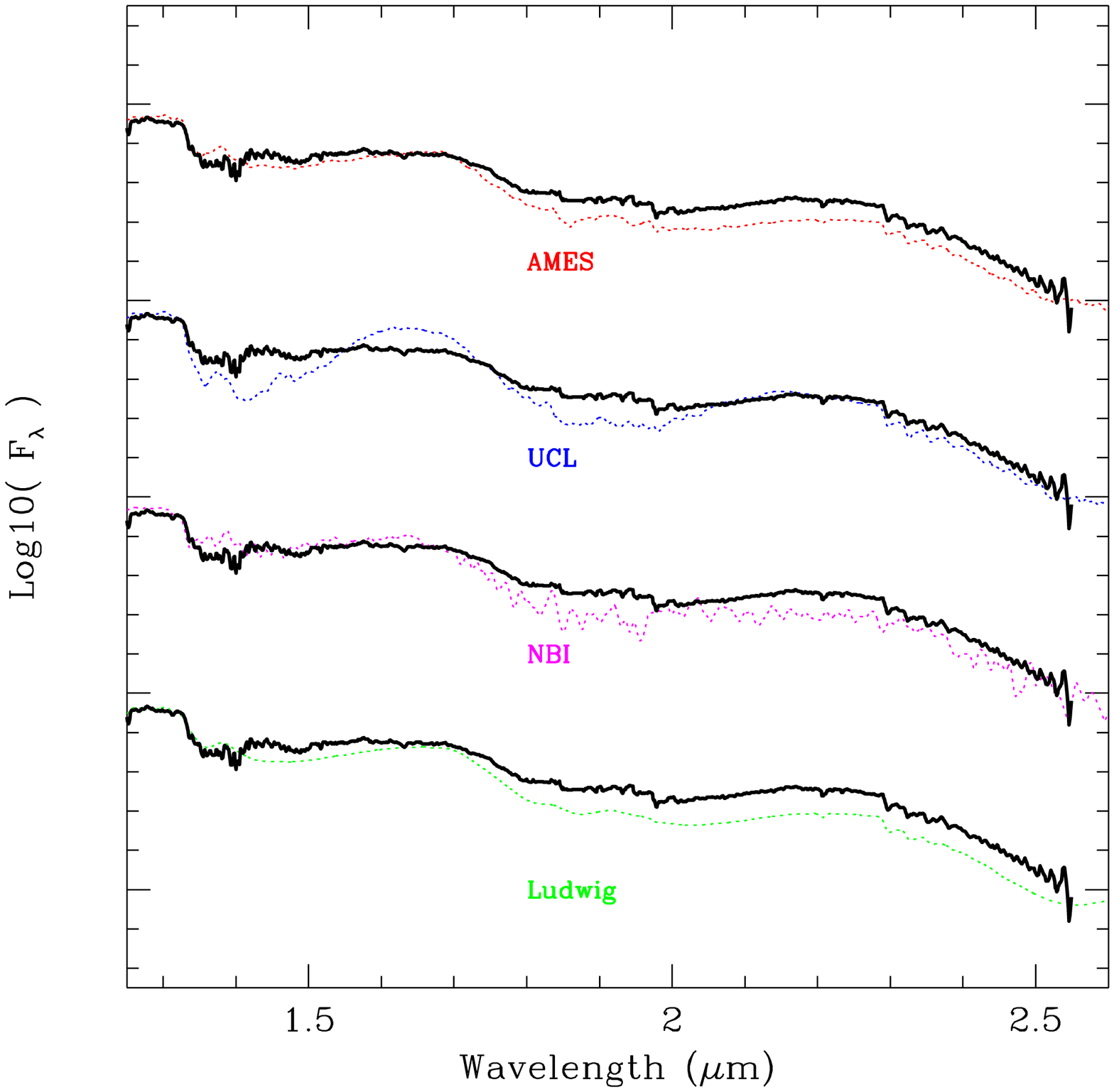}{8.5cm}{0}{54}{50}{-175}{-88}
  \caption{
    Synthetic spectra compared to the IR SED of VB10 using identical model
    parameters ($T_\mathrm{eff}=2800$\,K, $\mathrm{log}\,g=5.0$) and a 
    resolution of 50\,{\AA} with different water vapour opacity sources: the 
    Base grid by \cite{AH95} using \cite{LudwigH2O}; a test using the 1994 
    version of the Niels Bohr Institute \cite[]{NBIH2O}; the NextGen grid by 
    \cite{AllardH2O94,ARAA97} and \cite{NGa,NGb}  using the University 
    College London database \cite[]{UCL94H2O}; and the AMES-Cond/Dusty 
    grid by \cite{AllardTiOH2O2000,Allard01} using the NASA-Ames Center 
    database \cite[]{AMESH2O}. All models (except the NextGen/UCL case) 
    underestimate the flux at $K$ (ca.\ $2.0-2.4\,\mu m$) by 0.1 to 0.2 dex.}
  \label{f:allard_f1_water}
\end{figure}

\section{Molecular opacities}

While earlier work has been developed for the study of red giant
stars, the pioneering work on the modelling of VLM atmospheres has been
provided by \cite{Mould75}, \cite{AllardPhDT90} and \cite{PhDTKui91} using
a band model or Just Overlapping Line Approximation (JOLA) opacities
developed by \cite{Kivel52} and adapted for astrophysical use by
\cite{Golden67}.
More realistic model atmospheres and synthetic spectra for VLMs, brown
dwarfs and extrasolar planets have been made possible thanks to the
development of accurate opacities calculated often ab initio for 
atmospheric layers where temperatures can reach 3000\,K. 
The process of improvements was especially remarkable in the case of 
water vapour line lists.  Indeed, water vapour has seen an important 
evolution through the years from band model approximations to straight 
means based on hot flames experiments, and then to ab initio computations.  
Nevertheless, the atmosphere models have failed to reproduce the 
strength of the water bands that shape the low resolution
($R\le$\,300) infrared spectral energy distributions of M dwarfs. 
At the lower temperatures of brown dwarfs methane and ammonia rival
the effect of water. 
The discrepancies in the model synthetic spectra were therefore believed
to be due to inaccurate or incomplete molecular opacities.  In
particular water vapour was suspected because the 
discrepancies were observed at infrared wavelengths in the relative
brightnesses of the flux peaks between water vapour bands.  
As can be seen from Fig.\,\ref{f:allard_f1_water}
where the models are compared to the infrared spectrum of the M8e
dwarf VB10, the water vapour opacity profile which shape this part of
the spectrum has strongly changed over time with the improvement of
computational capacities and a better knowledge of the interaction
potential surface. And the most recent ab initio results confirm the earliest 
hot flames laboratory experiment results by \cite{LudwigH2O}. But in general, 
most opacity profiles produce an excess opacity (or lack of 
flux in the model) in the $K$ bandpass.  Only the UCL1994 line list
(due to incompleteness, and with much of its deviations cancelling out over 
the bandpasses) could produce seemingly correct $J-K_s$ colours.

\section{The revised solar abundances}

\begin{figure}[!ht] \plotone{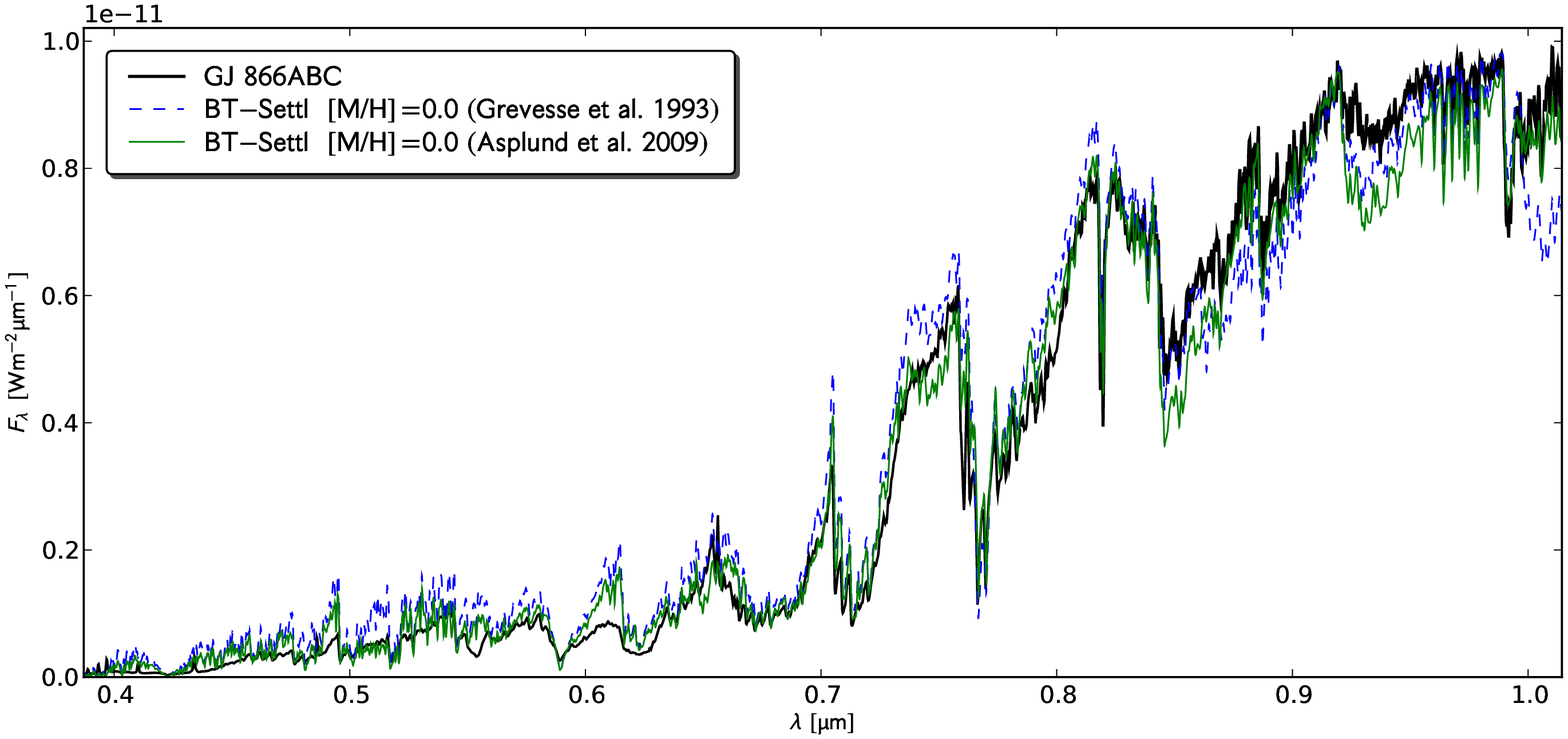}
  \plotone{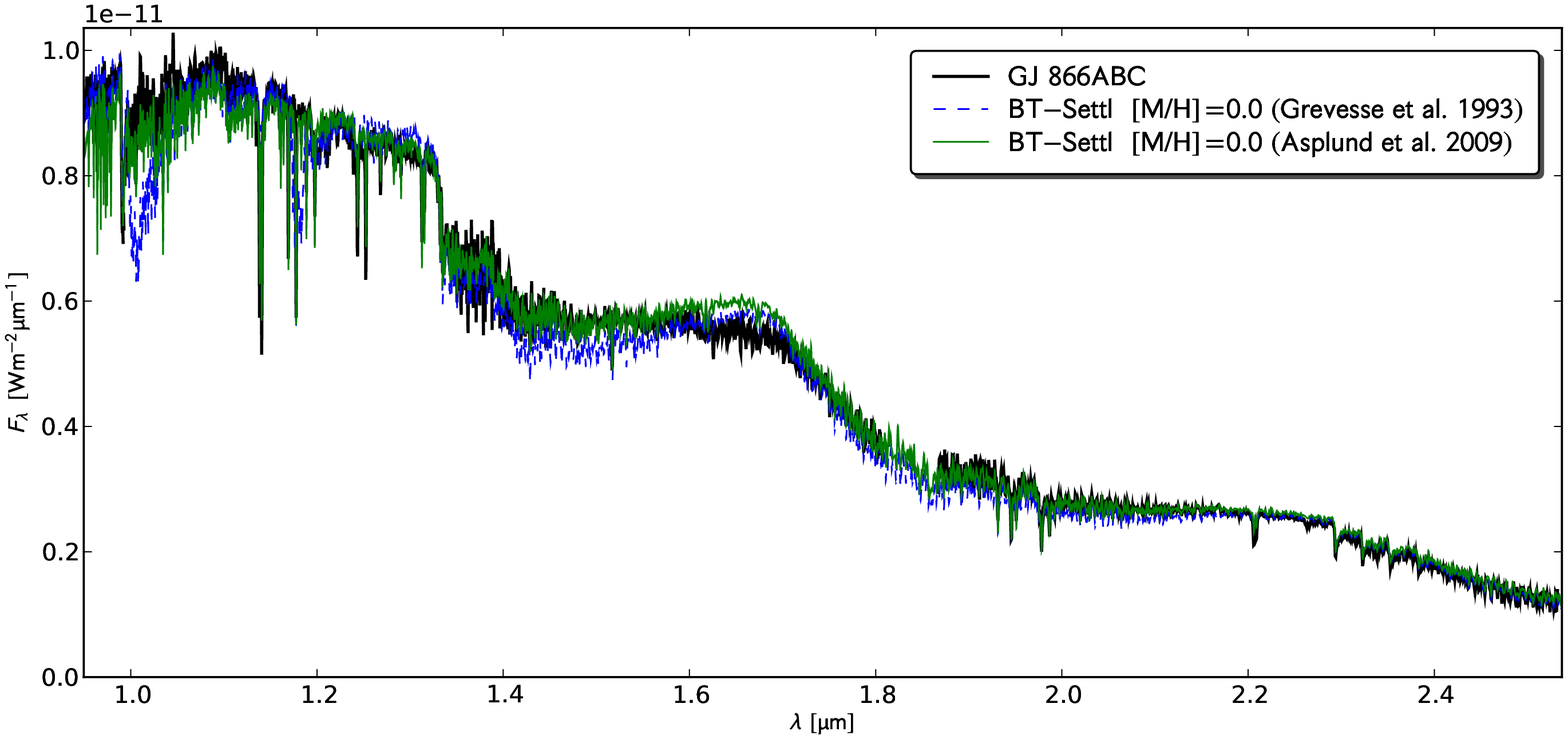}
  \caption{A BT-Settl synthetic spectrum with 
    $\mathrm{log}\,g$=5.0, and solar metallicity by \cite{Asplund09} is
    shown as light solid line, compared to the combined SED of the red
    dwarf triple system Gl\,866 \citep{Leinert1990,Delfosse1999}. 
    The observations were combined from a Mt.\ Stromlo optical spectrum 
    (M. Bessell, priv.\ comm.) and SpeX infrared spectrum taken at the 
    NASA IRTF \citep{IRTF2009} (thick curve). 
    For comparison a model using the same parameters and physical setup
    with the \citet{GNS93} abundances is shown as a dashed line. 
    The models have been scaled to the observed absolute flux assuming two
    equal $T_\mathrm{eff}$=2920\,K components of 0.157 solar radii and a
    third with $T_\mathrm{eff}$=2700\,K and 0.126\,$R_\odot$.}
  \label{f:allard_f2_GJ866}
\end{figure}

\begin{figure}[!ht] 
  \plotfiddle{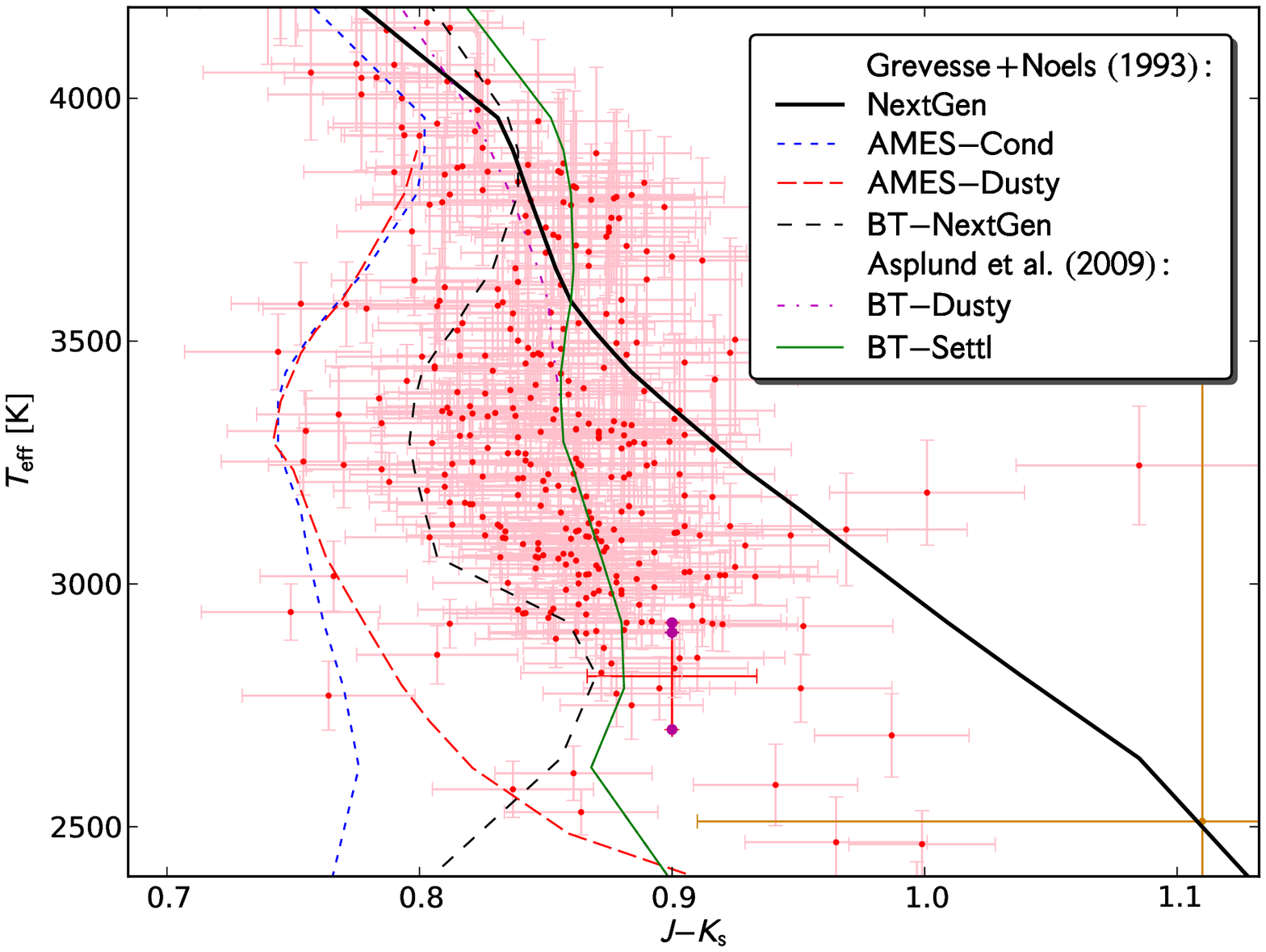}{7.0cm}{0}{55}{55}{-170}{-125}
  \caption{Estimated $T_\mathrm{eff}$ for M dwarfs by \cite{MdwarfsTeff2008} 
    and brown dwarfs by \cite{Golimowski04,Vrba04} are compared to the NextGen 
    isochrones for 5 Gyrs  \cite[]{BCAH97,BCAH98} using various generations of 
    model atmospheres:  NextGen (thick line), the limiting case AMES-Cond/Dusty 
    cases by \cite{Allard01} (dotted and dashed lines), the current BT-Settl models 
    using the \cite{Asplund09} solar abundances (full line).  The Gl\,866 system fitted 
    in Fig.\,\ref{f:allard_f2_GJ866} is highlighted by darker colours and shown with its 
    relatively large photometric error bars at $J-K_\mathrm{s} = 0.9$.}
  \label{f:allard_f3_Teff-J-Ks}
\end{figure}

Model atmospheres for VLMs and in general for other stars assume 
scaled solar abundances for all heavy elements, with some enrichment
of $\alpha$-process elements (the result of a "pollution" of the
star-forming gas by the explosion of a supernova)  
when appropriate in the case of metal-depleted subdwarfs
of the Galactic thick disk, halo and globular clusters.  The revision of the solar
abundances based on radiation hydrodynamical simulations of the solar
atmosphere, on improvements in the quality of the spectroscopic
observations of the Sun, and in its detailed line profile analysis by two
separate groups using independent hydro codes and
spectral synthesis codes \citep[]{Asplund09,Caffau2011} yield an 
oxygen reduction of 0.11\,--\,0.19~dex (up to 34\%).  
compared to the previously used abundances of \citet{GNS93}. 
Since the overall SED of late K dwarfs, M dwarfs, brown
dwarfs, and exoplanets is governed by oxygen compounds (TiO, VO in the
optical and water vapour and CO in the infrared), the input elemental oxygen
abundance used in the equation of state is of major importance.
Fig.\,\ref{f:allard_f2_GJ866} shows an example of these effects for the
optical and infrared SED of the M5.5 dwarf system Gl\,866. 
However at other effective temperatures even stronger photometric
effects can be seen, where the near-IR SED of different models
diverges more (see Fig.\,\ref{f:allard_f3_Teff-J-Ks}).  
The comparison shows significant improvement compared to older 
models shown in Fig.\,\ref{f:allard_f1_water},  except for excess flux 
in the $H$ bandpass near 1.7\,$\mu m$ due to incomplete FeH opacity 
data for this region.  The comparison has particularly improved in the 
Wing Ford band of FeH near $0.99~{\mu}$m, and in the VO bands 
thanks to line lists provided by B. Plez (GRAAL, Montpelier, France), although 
inaccurate or incomplete opacities are still affecting the models at optical wavelengths  
\citep[e.g. the TiO line list by][]{AMESTiO}. 

Fig.\,\ref{f:allard_f3_Teff-J-Ks} compares the theoretical isochrones
(assuming an age of 5 Gyrs) to the \cite{MdwarfsTeff2008}  $T_\mathrm{eff}$
estimates and reveals that the NextGen models 
\cite[]{ARAA97,NGa,NGb} systematically and  increasingly overestimate 
$T_\mathrm{eff}$ through the lower main sequence, while the 
AMES-Cond/Dusty \cite[]{Allard01} models on the contrary underestimate 
$T_\mathrm{eff}$ as a function of $J-K_\mathrm{s}$ colour. This
situation is relieved when using the current models 
(labeled BT-Settl in the figure) based on the revised solar abundances, and 
the models now agree fairly well with most of the empirical estimations of 
$T_\mathrm{eff}$. The current model atmospheres have not yet been used 
as surface boundary condition to interior and evolution calculations, and 
simply provide the synthetic colour tables interpolated on the
published theoretical isochrones \cite[]{BCAH98}. 
Even if the atmospheres partly control the cooling and evolution of M
dwarfs \cite[]{CB97}, differences introduced in the surface boundary
conditions by changes in the model atmosphere composition have 
negligible effect. 

\section{Cloud formation}

One of the most important challenges in modelling these atmospheres
(below 2600K) is the formation of clouds. \cite{Tsuji96a} had identified
dust formation by recognising the condensation temperatures of hot 
dust grains (enstatite, forsterite, corundum: MgSiO$_3$, Mg$_2$SiO$_4$, 
and Al$_2$O$_3$ crystals) to occur in the line-forming layers 
($\tau \approx 10^{-4} - 10^{-2}$)  of their atmospheres. The cloud composition, 
according to equilibrium chemistry, is going from zirconium oxide (ZrO$_2$), 
to refractory ceramics (perovskite and corundum; CaTiO$_3$, Al$_2$O$_3$), 
to silicates (e.g. forsterite; Mg$_2$SiO$_4$), to salts (CsCl, RbCl, NaCl), and 
finally to ices (H$_2$O, NH$_3$, NH$_4$SH) as brown dwarfs cool 
down over time from M through L, T, and Y spectral types \cite[]{Allard01,LF06}.
This assumed (by Allard et al. 2001) sub-micron-sized crystal formation causes the weakening
and vanishing of TiO and VO molecular bands (via CaTiO$_3$, TiO$_2$,
and VO$_2$ grains) from the optical spectra of late M and L dwarfs,
revealing CrH and FeH bands otherwise hidden by the molecular
pseudo-continuum, and the resonance doublets of alkali transitions
which are only condensing onto salts in late-T dwarfs. The scattering
effects of this fine dust is Rayleigh scattering which provides veiling to
the optical SED of late-M and L dwarfs, while the greenhouse effect
due to the dust cloud causes their infrared colours to become
extremely red compared to those of hotter low mass stars.  The upper
atmosphere, above the cloud layers, is depleted from condensible
material and significantly cooled down by the reduced or missing
pseudo-continuum opacities.

\begin{figure}[!ht] 
  \plotfiddle{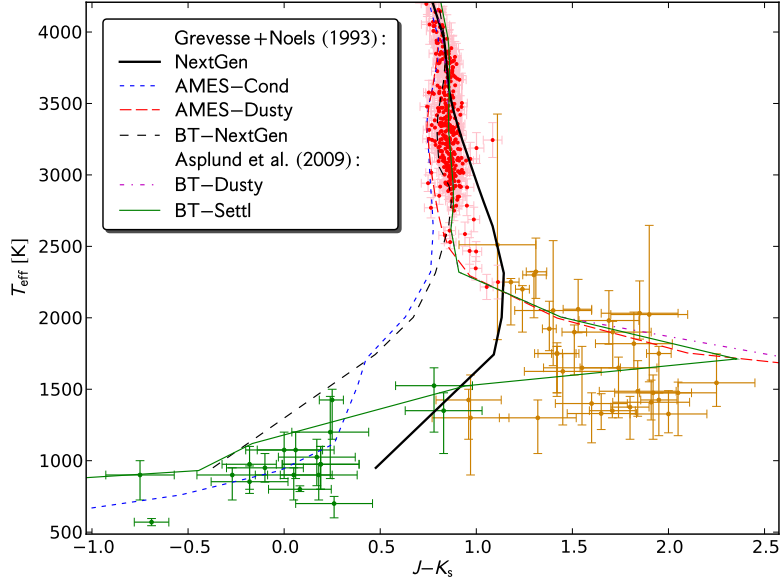}{7.0cm}{0}{55}{55}{-170}{-125}
\caption{Same plot as Fig.\,\ref{f:allard_f3_Teff-J-Ks} but zooming out
and extending into the brown dwarf region of the diagram. This region
below 2500K is dominated by dust formation (essentially forsterite and
other silicates).  The limiting cases AMES-Cond/Dusty model
atmospheres provide a description of the span in colours of the brown
dwarfs in this diagram for a given age (here 5 Gyrs). The BT-Settl models
succeed in explaining even the most extreme colours of brown dwarfs.}
\label{f:allard_f5_Teff-J-Ks_cool}
\end{figure}

One common approach has been to explore the limiting properties of 
cloud formation. One limit is the case where sedimentation or gravitational settling 
is assumed to be fully efficient such as the case B of \cite{Tsuji02}, the 
AMES-Cond or condensed phase models of \cite{Allard01}, the clear case 
of \cite{AM01} and the cloud-free case of \cite{Burrows06}. 
The other limit is the case where gravitational settling is assumed inefficient and dust, 
often only forsterite, forms in equilibrium with the gas phase such as the
case A of \cite{Tsuji02}, the AMES-Dusty or dusty models of \cite{Allard01}, 
the cloudy case of \cite{AM01}, or the case B of \cite{Burrows06}. 
These limiting cases of maximum dust content agree in describing the 
evolution of brown dwarfs from a molecular opacity governed SED 
towards a blackbody SED below 1500K. This description was 
suitable, at least in the case of the AMES-Dusty models, in reproducing 
the infrared colours of L dwarfs.
The cloud-free limiting case on the other hand allowed to reproduce to 
some degree the colours of T dwarfs.  Fig.\,\ref{f:allard_f5_Teff-J-Ks_cool} 
shows this situation for the AMES-Cond/Dusty limiting case models
of \cite{Allard01} compared with the effective temperatures estimates
obtained by integration of the observed SED \cite[]{Golimowski04,Vrba04}.

The purpose of a cloud model is therefore to go beyond these limiting
cases and define the number density and size distribution of
condensates as a function of depth in the atmosphere. The discovery of
dust clouds in M dwarfs and brown dwarfs has therefore triggered the
development of cloud models building up on pioneering work 
in the context of planetary atmospheres developed by
\cite{Lewis69}, \cite{Rossow78}, and \cite{Lunine89}.  The Lewis model
is an updraft model (considering that condensation occurs in a gas
bubble that is advected from deeper layers). By lack of knowledge of
the velocity field and diffusion coefficient of condensates in the
atmospheres of the planets of the solar system, Lewis simply assumed
that the advection velocity is equal to the sedimentation velocity,
thereby preserving condensible material in the condensation
layers. This cloud model did not account for grain sizes. Rossow on the
other hand developed characteristic timescales as a function of
particle size for the main microphysical processes involved
(condensation, coagulation, coalescence, and sedimentation). The
intersections of these characteristic timescales gives an estimate of
the condensate number densities and mean grain sizes. However, this
model made several explicit assumptions concerning the efficiency of
supersaturation, the coagulation, etc.  

\cite{Helling08b} have compared different cloud models and their impact 
on model atmospheres.  Most cloud models define the cloud base as the 
evaporation layer provided by the equilibrium
chemistry. In the unified cloud models of \citet{Tsuji02,tsujiCloudII} 
a parametrization of the radial location of the cloud top by way of an
adjustable parameter T$_{\rm crit}$ was used. This choice permits to
determine the cloud extension 
effects on the spectra of these objects but does not allow to reproduce the 
stellar-substellar transition with a unique value of T$_{\rm crit}$ since the 
cloud extension depends on the atmospheric parameters.

\cite{Allard03} using \texttt{PHOENIX} and the index of refraction of up to 40
condensible species, have applied the Rossow cloud model, ignoring
coalescence and coagulation, and comparing the timescales of
condensation, sedimentation and mixing (extrapolated from the
convective velocities into the convectively stable layers), and
assuming efficient nucleation (monomers equilibrium densities). The
cloud model was then solved layer by layer inside out to account for
the sequence of grain species formation as a function of cooling of
the gas. But this version of the BT-Settl (with gravitational settling) models 
did not allow for the formation of enough dust in brown dwarf atmospheres 
due to a too conservative prescribed supersaturation value. 

\cite{AM01} have solved the particle diffusion problem of
condensates assuming a parametrized sedimentation efficiency $f_{\rm
sed}$ (constant through the atmosphere) and a mixing assumed constant
and fixed to its maximum value (maximum of the inner convection
zone). \citet{SaumonMarley2008} found that their models could not 
produce the color change with a single value of  $f_{\rm sed}$.

\cite{Helling08a} use the \texttt{PHOENIX} code to compute the Drift-Phoenix
models. The cloud model used, in the contrary to all other cases 
mentioned, studies the nucleation and growth of grains as they sediment 
down into the atmosphere. This cloud model determines the number 
density and size distribution of grains by 1D nucleation simulations, and 
the resulting distribution is read in by \texttt{PHOENIX} which computes 
the resulting opacities and radiative transfer.  These models solve 
the nucleation problem but only for the assumed monomer types and 
have been successfully applied to fit the dusty atmospheres of L dwarfs,
but the reversal in IR colours observed for the L/T transition could not be 
explained \citep{WitteDRIFT_III}. 



None of these models however treated the mixing properties of the atmosphere 
and the resulting  diffusion mechanism realistically enough to
reproduce the brown dwarf  spectral transition without changing cloud
parameters.  
\cite{Freytag2010} have therefore addressed the complementary though
important issue of mixing and diffusion in these atmospheres by 2D
radiation hydrodynamic (hereafter RHD) simulations, using the 
\texttt{PHOENIX} gas opacities in a multi-group opacity scheme, and
forsterite with geometric cross-sections. These simulations assume
efficient nucleation, using monomer densities estimated from the total
available density of silicon (least abundant element in the solar
composition involved in forsterite).  They found that gravity waves
play a decisive role in clouds formation, while around $T_\mathrm{eff}
\le$~2200\,K the cloud layers become optically thick enough to
initiate cloud convection, which participate in the 
mixing. Overshoot can also be important in the deepest layers.  

These RHD simulations allow an estimation of the diffusion 
processes bringing fresh condensible material from the hotter lower layers to
the cloud forming layers.  We have therefore updated our cloud model
(BT-Settl models) to account for the mixing prescribed by 
the RHD simulations. Another important improvement concerns the supersaturation 
which as been computed rather than using the fixed conservative value 
recommended by Rossow\nocite{Rossow78}. One can see from
Fig.\,\ref{f:allard_f5_Teff-J-Ks_cool} that the late-type M and
early-type L dwarfs behave as if dust is formed nearly in equilibrium
with the gas phase with extremely red colours in some agreement with the
BT-Dusty models.  The BT-Settl models reproduce the main sequence down to
the L-type brown dwarf regime, subjected in the $K$ bandpass to the 
greenhouse effect of dust clouds, before turning to the blue in the
late-L and T dwarf regime as a result of methane formation in the $K$
bandpass. 
This constitutes a major improvement over previous
models, and is promising that we can reach in the near futur a full
explanation of the stellar substellar transition.

Diffusion has also been held responsible for deviations in ultracool
atmospheres from gas phase chemical equilibrium (CE), as noted in
early observations of T dwarfs showing an excess of carbon monoxide
absorption \citep{noll97Gl229B,gyCO99}. More recently, carbon
dioxide \citep{TsujiAKARI}, which was not expected at such low temperatures, has been detected. 
Similarly, ammonia has been shown to be underabundant
\citep{didierGl570}. This is understood as the result of slowing down
of crucial chemical reaction steps, 
so that some important molecules (CH$_4$, NH$_3$) would not have the
time to form in equilibrium while undergoing mixing, whereas others
(CO, CO$_2$, N$_2$) remain at enhanced abundances. 
The RHD simulations of \cite{Freytag2010} have allowed to understand
the underlying mixing processes, obviating the need to describe them
with an additional free parameter. 

\section{Applications to Exoplanet Science}

Several infrared integral field spectrographs combined with
coronograph and adaptive optic instruments being developed are
coming online before 2013 (SPHERE at the VLT, the Gemini Planet Imager
at Gemini south, Project1640 at Mount Palomar, etc.). The E-ELT
41m telescope in Spain due around 2020 will also be very ideally
suited for planet imaging.  The models developed for VLMs and brown
dwarfs are a unique opportunity, if they can explain the stellar-substellar 
transition, to provide a great support for the characterisation of imaged
exoplanets.  We have therefore developed the BT-Settl model atmosphere 
grid to encompass the parameter regime of these objects (surface gravity 
around $\mathrm{log}\,g$=4.0, $T_\mathrm{eff}$ below 2000K). 

These planets are typically found at several dozens of AU
from the star, and since the observations are done in the infrared the
non-irradiated models can even be used directly. Indeed,
\cite{Barman2001} have shown that the effects of impinging radiation
from a star on the planetary atmosphere are Rayleigh scattering of the
stellar light by H$_2$ molecules (or clouds if present) at optical
wavelengths (below 1 $\mu$m for solar type stars), while the impact on
the interior and evolution properties becomes negligible for orbital
distances exceeding 0.1 AU. Nevertheless, we are developing for 2012 
irradiated models and the capacity to compute them via the \texttt{PHOENIX} 
simulator (see below).

\section{Summary and Futur Prospects}

We report progress of the development of a new model atmosphere grid
for stars, brown dwarfs and young planets, named BT-Settl. It has been
computed using the \texttt{PHOENIX} code updated for: i)
the \cite{BT2H2O} BT2 water, the \cite{stdsIAU211} STDS methane, the
\cite{Sharp2007} ammonia and the \cite{cdsd1000} CDSD-1000 CO$_2$
opacity line lists, ii) the solar 
abundances revised by \cite{Asplund09}, and iii) a cloud model
accounting for more detailed supersaturation and RHD mixing.  The grid
is covering the whole range of stars to young planets 400\,K $<
T_\mathrm{eff} <$ 70,000\,K; -0.5 $<\mathrm{log}\,g<$ 5.5; and $-4.0
<$ [M/H] $< +0.5$, including values of the $\alpha$-element
enhancement   (supernovae enrichment of the star forming material) 
between +0.0 and +0.6.
Models are available at the \texttt{PHOENIX} simulator website
\texttt{http://phoenix.ens-lyon.fr/simulator/} and are in preparation
for publication to serve among others the GAIA, MUSE and
SPHERE/GPI/P1640 instruments to come online in the near futur. 
Corresponding evolution models are expected for 2012.

We found the previously used NextGen models to systematically
overestimate $T_\mathrm{eff}$ below 3500\,K by as much as 500\,K.  The
water vapour opacity profile has converged with the most recent line
lists reproducing laboratory results, but could not explain this
discrepancy. The solution came instead from the revision of the solar
abundances which changes the strengh of the water vapour absorption
bands, and therefore allows the reproduction of the spectroscopic and
photometric properties of M dwarfs as late as M6. Later-type M dwarfs
are affected by dust formation and cloud modelling is important to
understand their properties.  We find that the Rossow cloud model
allows, with revisions to the supersaturation and mixing, the
reproduction of the stellar-substellar transition.  A small offset
persists however in the M-L transition.  It is possible that all the
current cloud models are not efficient enough in producing dust at the
onset of cloud formation regime.  Detailed nucleation studies could
allow in the futur to resolve this issue. Other uncertainties affect
the current cloud modelling such as the assumption of spherical
non-porous grains while grains form as fractals in the
laboratory. Constraining the models remains therefore very important.

Beyond cloud modelling and molecular opacities, model atmospheres for
these objects require reaction rates for the most abundant molecules
and/or most important absorbers.  Furthermore, these atmospheres are
composed of molecular hydrogen which constitute the main source of
collisions.  Also needed are therefore collision rates (by H$_2$) and
corresponding damping constants for the broadening molecular lines.

In order to say something about the spectral variability of VLMs,
brown dwarfs and planets, 3D global or "star in a box" RHD simulations
with rotation will be required.  This is our current project supported
by the French ``Agence Nationale de la Recherche'' for the period
2010-2015.

\acknowledgements We thank the French ``Agence Nationale de la
Recherche'' (ANR), and the ``Programme National de Physique
Stellaire'' (PNPS) of CNRS (INSU) for their financial support. The
computations of dusty M dwarf and brown dwarf models were performed at
the {\sl P\^ole Scientifique de Mod\'elisation Num\'erique} (PSMN) at
the {\sl \'Ecole Normale Sup\'erieure} (ENS) in Lyon and at the {\sl
Gesellschaft f{\"u}r Wissenschaftliche Datenverarbeitung
G{\"o}ttingen} in collaboration with the Institut f{\"u}r Astrophysik
G{\"o}ttingen.
\bibliography{allard_f}

\end{document}